%% file: CDC2020_ExtendedFBSP_Modif_EV.tex
\documentclass[letterpaper, 10 pt, conference]{ieeeconf}  

\IEEEoverridecommandlockouts                              

\usepackage{comment}
\usepackage{soul}

\usepackage{xcolor}
\usepackage{psfrag}
\usepackage{pgfplots}

\usepackage{mathrsfs}
\usepackage{amsbsy}

\usepackage{graphics} 
\usepackage{epsfig} 
\usepackage{mathptmx} 
\usepackage{times} 
\usepackage{amsmath} 
\usepackage{amssymb}  
\usepackage{mathtools}
\usepackage{array}
\usepackage{bm}
\usepackage{tikz}
\usetikzlibrary{calc,arrows}
\usepackage{pgfplots}
\usepackage{subcaption}
\usepackage{etex}
\usepackage{siunitx}
\usepackage{url}
\usepackage{arydshln} 
\usepackage[]{algorithm2e, setspace}
 \let\labelindent\relax
\usepackage{enumitem} 

\usepackage[noend]{algpseudocode}

\makeatletter
\def\BState{\State\hskip-\ALG@thistlm}
\makeatother

\DeclareMathOperator*{\diagonal}{diag}

\usepackage[backend=bibtex, style=ieee,doi=false,
    isbn=false,
    url=false,
    eprint=false,
]{biblatex}
\bibliography{bibliography}

\title{Extended Full Block S-Procedure for Distributed Control of Interconnected Systems*}

\author{Giulia De Pasquale$^{1}$, Yvonne R.\ St\"urz$^{2}$, Maria Elena Valcher$^{1}$ and Roy S.\ Smith$^{3}$ 
\thanks{$^{1}$The authors are with the Dipartimento di Ingegneria dell'Informazione Università di Padova, Via Gradenigo 6b, 35131 Padova, Italy 
{\tt\small \{giulia.depasquale,meme\}@dei.unipd.it }
}
\thanks{$^{2}$The author is with the Model Predictive Control Laboratory, University of California, Berkeley, CA 94709, USA. Email address: %
        {\tt\small y.stuerz@berkeley.edu }}
 \thanks{$^{3}$The author is with the Automatic Control Laboratory, ETH Zurich, Physikstrasse 3, 8092 Zurich, Switzerland     {\tt\small rsmith@control.ee.ethz.ch}}
\thanks{*This project has received funding from the European Union’s Horizon 2020 research and innovation programme under the Marie Sklodowska-Curie grant agreement No.\ 846421.
Paper accepted at the 59th IEEE Conference on Decision and Control (CDC), Jeju Island, Korea.}}%

\include{nomenclature}

\begin{document}

\maketitle
\thispagestyle{empty}
\pagestyle{empty}


\begin{abstract}
This paper proposes a novel method for distributed controller synthesis of homogeneous interconnected systems consisting of identical subsystems. 
The objective of the designed controller is to minimize the $\mathcal{L}_2$-gain of the performance channel. 
The proposed method is an extended formulation of the Full Block S-Procedure (FBSP) where we introduce an additional set of variables. This allows to relax the block-diagonal structural assumptions on the Lyapunov and multiplier matrices required for distributed control design, which reduces conservatism w.r.t.\ most existing approaches. 
We show how to decompose the proposed extended FBSP into small synthesis conditions, of the size of one  individual subsystem. 
%
%
\end{abstract}

\section{Introduction}
A wide variety of modern control applications involve complex systems that are characterized by their large scale, their distributed nature and the sparse structure of their physical interconnections \cite{Rantzer1}. Examples include smart grids, the internet, automated highways \cite{Rantzer1}, satellite formation flying  \cite{satellite}, car platoons \cite{platoons}, cooperative manipulation \cite{Stuerz2017a}, unmanned aerial vehicles \cite{unmanned2} and large telescopes \cite{massioni2010}. This gave rise to an increased interest in the research on modelling and control of complex distributed systems, consisting of  many subsystems interacting through a network. 
Distributed systems represent a class of interconnected systems. 
When systems are physically interconnected we refer to them as "spatially distributed systems" and this is the case of automated highway systems \cite{highway}, airplane formation flight, satellite constellations \cite{satellite}. 
We call, instead,  "virtually interconnected systems" those systems whose subsystems are not spatially interconnected, but share information in order to reach a common goal, such as mobile robots \cite{robots}.  


The control of large-scale networked systems is a challenging problem since a large number of inputs and outputs are involved. 
Because of limitations in computation and communication, the synthesis and the implementation of a centralized controller becomes intractable in practice, and decentralized or distributed control architectures are required. However, in applications with strong interactions among the subsystems, a completely decentralized architecture of the controller may not guarantee good performance  \cite{Stuerz2017b}. 
With a growing number of subsystems, the controller synthesis becomes more complex. Therefore, decomposition methods have been proposed for a scalable controller synthesis. 
To this end, the system can be modeled as consisting of  a decentralized part and an interconnection part. 
This is achieved by introducing an interconnection channel, that captures the interconnections among subsystems, and by performing a linear fractional transformation (LFT) on the system model   \cite{Langbort2004}. 
Groups of identical interconnected subsystems  constitute a homogeneous system \cite{massioni2010}. This model has been extended to groups of homogeneous systems \cite{massioni3},  
and groups of homogeneous interconnections \cite{Stuerz2020, Stuerz}. 
For groups of homogeneous systems, the system structure can be exploited in order to derive a compact controller synthesis. 
In \cite{massioni2010,massioni3,Stuerz}, the Full Block S-Procedure (FBSP) has been exploited for the synthesis of a distributed controller, resulting in linear matrix inequalities (LMIs). The decomposition and thus scalability of the synthesis conditions are achieved by a block-diagonal structure of the Lyapunov and multiplier matrices, which introduces conservatism. 
In this paper, we propose a novel distributed controller synthesis for linear time invariant homogeneous interconnected systems, which allows to reduce this conservatism. 
This is achieved by proposing an extended version of the FBSP which involves an additional set of variables. This allows us to relax the imposed structure on the Lyapunov and multiplier matrices, and can thus lead to a better control performance. 
A similar approach has been proposed in \cite{Farhoodi2008} with an extended version of the Bounded Real Lemma. 
The extended FBSP (EFBSP) proposed in this paper can be decomposed to allow for a scalable distributed controller synthesis. We show that by decomposing the EFBSP we achieve the same degree of scalability w.r.t.\ the FBSP with block-diagonal structured Lyapunov and multiplier matrices, namely a decomposition into small matrix inequalities which are of the size of the individual subsystems is achieved. 

The paper is structured as follows. 
In Section~\ref{subsec:Graphstructure} interconnected systems are introduced with special focus on homogeneous systems.  Section~\ref{sec:problem_formilation} formalizes the synthesis problem of the interconnected controller. In Sections~\ref{sec:prima_FBSP} and \ref{sec:dual_EBRP} the primal and dual formulations of the EFBSP are given, while in Section~\ref{sec:EFBSP_decomp} the decomposition of the controller synthesis equations based on the EFBSP is introduced.  Section~\ref{sec:sim_res} shows some numerical results. 

\textbf{Notation} We denote a block-diagonal matrix $D$ with diagonal blocks $D_1,...,D_N$ by $D = \diagIndtwo{i=1}{N}{D_h}$. The $n\times n$-identity matrix is denoted by $I_n$ and the $n \times m$ matrix of all zeros by $0_{n \times m}$. If clear from the context, the indices are dropped. 
Given a matrix $M$, its minimum and maximum singular values are denoted by $\sigma_{\mathrm{min}}(\cdot)$ and $\sigma_{\mathrm{max}}(\cdot)$, respectively, while $spec(M)$ denotes its spectrum, i.e. the set of all its eigenvalues. 
The Kronecker product is denoted by $\kronecker{}{}$.
Given a complex valued matrix $M=\small \bma{@{\,}c@{\,\;}c@{\,}}{M_1 & M_2\\ M_3 & M_4}$ and a matrix $P$ of appropriate dimensions,   the lower and upper Linear Fractional Transformation (LFT) associated with the pair $(M,P)$ are defined as $\mathcal{F}_\ell(M,P) =M_1+M_2P(I-M_4)^{-1}M_3$ and $\mathcal{F}_\text{u}(M,P) =M_4+M_3P(I-M_1)^{-1}M_2$, respectively. We use the symbol $\star$ to simplify expressions as $M_1^\top M_2 M_1$, i.e., $\star^\top M_2 M_1 = M_1^\top M_2 M_1$.

\section{Homogeneous Interconnected Systems}\label{subsec:Graphstructure}
In this paper, we consider homogeneous interconnected systems which are composed of $N$ identical subsystems. The subsystems can be coupled through their dynamics, disturbance, performance goal or communication. 
 Assuming that each subsystem, denoted by $G_h$, $h \in \mathcal{N}$, $\mathcal{N}:= \{1,\dots,N\}$, is of order $n_h${\color{black}$=n$}, the state-space dynamics of the whole system, of order $n_x:=\sum_{h \in \mathcal{N}}n_h${\color{black} = $Nn$}, is described as 
\begin{equation}\label{eq:origsys_matrices}
\begin{cases}
\dot{x}(t) &= Ax(t)+B_ww(t)+B_uu(t)\\ 
y(t) &= C_yx(t)+D_{yw}w(t) \\
z(t) &= C_zx(t)+D_{zw}w(t)+D_{zu}u(t)\end{cases}
\end{equation}
with the state vector $x \in \mathbb{R}^{n_x}$, 
the control input and measured output, $u \in \mathbb{R}^{n_{u}}$ and $y \in \mathbb{R}^{n_{y}}$, the exogenous input and performance output, $\perfi \in \mathbb{R}^{n_{\perfi}}$ and $\perfo \in \mathbb{R}^{n_{\perfo}}$.

We define the undirected and unweighted graph that represents the coupling among the subsystems (in terms of dynamics, performance or disturbance) as $\mathcal{G} = \{\mathcal{N},\mathcal{E}\}$, where the index set $\mathcal{N}$ represents the $N$ subsystems $G_h$, and $\mathcal{E}\subseteq \mathcal{N}\times \mathcal{N}$ represents the interconnections between the subsystems. In particular, the pair $(h,j)$ belongs to $\mathcal{E}$, for $h \neq j$,  if and only if subsystems $G_j$  and $G_h$ communicate.
The interconnections among the subsystems $G_h$ are described by a pattern matrix denoted  by $P$:
  its  $(h,j)$th entry  is $1$ if and only if  $(h,j)\in \mathcal{E}$, otherwise it is zero. We assume that ${\mathcal G}$ has no self loops and hence the diagonal entries of $P$ are zero. The pattern matrix $P$ fully characterizes the topology of the interactions among the subsystems. For simplicity, we have considered undirected graphs, i.e., we have assumed that the pattern matrix $\pat$ is symmetric.
This ensures that the eigenvalues of $\pat$ are real. 
%
However, the present results can be extended to directed graphs as well, as discussed in \cite{massioni2}. 

\subsection{Homogeneous Interconnected State-Space Representation}
\label{subsec: interconnected_sys}

For a homogeneous interconnected system, the system matrices in \eqref{eq:origsys_matrices} have the following structure: 
\begin{mydef}[Homogeneous Decomposable System]\label{def:homo_sys}
	Let $M$ generically represent one of the system matrices $A$, $B_w$, $B_u$, $C_y$, $D_{y \perfi}$, $C_{\perfo}$, $D_{\perfo u}$, $D_{\perfo \perfi}$, in \eqref{eq:origsys_matrices}. 
 System  \eqref{eq:origsys_matrices} is said to be a homogeneous decomposable system if each of its system matrices can be written as 
\begin{equation}
		\label{eq:M_homo}
		M =  \kronecker{I_N}{ {M^d}}  + 
		 \kronecker{ {\pat} }{ M^i},   
	\end{equation}
	with ${\pat}$ being the pattern matrix  defined before, for two suitable matrices $M^d$ and $M^i$. 
	This means that for a homogeneous system all local subsystem matrices (on the diagonal) are identical and equal to $M^d$, and  all interconnection subsystem matrices  are equal to $M^i$
	(the off-diagonal block $(h,k), h\ne k,$ is equal to $M^i$ if $(h,k)\in {\mathcal E}$, the zero matrix otherwise). 
\end{mydef}

Adopting an approach similar to \cite{massioni3}, the system can be  modeled through an LFT of a decentralized part and an interconnection part, as a result of the introduction of an interconnection channel. The diagonal-blocks of the system matrices \eqref{eq:origsys_matrices} model the decentralized part and the off-diagonal ones refer to the interconnection part.
 We introduce the interconnection channel ${q \rightarrow p}$ as 
 \begin{equation}\label{eq:interconn}
 {p= \mathcal{P} q},
 \end{equation}
 where {\color{black} $\intercPi \in \mathbb{R}^{n_{\intercPi}}$ and $\intercPo \in \mathbb{R}^{n_{\intercPo}}$ are the interconnection variables.} {\color{black} Specifically, the $h$-th block of $p$ represents the interconnection variables available as inputs to the $h$-th subsystem, while the $j$-th block of  $q$ represents the variables of subsystem $G_j$ that are accessible to those subsystems $G_h$ for which} $(h,j)\in \mathcal{E}$. 
 The interconnection matrix $\mathcal{P}$ in \eqref{eq:interconn} captures the interconnection relations, hence the system can be modeled by the LFT $\mathcal{F}_\ell(G,\mathcal{P})$, with $G = I_N \otimes G_h$, where $G_h$ are the decentralized parts of the subsystems. For ideal  deterministic interconnections considered in this work $\mathcal{P}$ can be defined as $\mathcal{P} = \kronecker{P}{I_{n_{\intercPo}}}$.

The system vector in \eqref{eq:origsys_matrices} is structured as $x = [x_1^\top, ..., x_N^\top]^\top$, and analogously  the input and output vectors, and the interconnection variables in \eqref{eq:interconn}. 
The $N$ subsystems $G_h$ of the system in  \eqref{eq:origsys_matrices} then admit a continuous-time state-space representations given by   

\begin{equation}
	\label{eq:interconn_openloop_h}
	\begin{aligned}
		\Plant_h & : 
		\begin{cases}
			\bma{@{}c@{}}{ 
				\dotxP_h \\
				\hdashline 
				y_h\\
				\hdashline 
				\perfo_h \\
				\hdashline
				\intercPo_h \\
			} 
			= 
			\bma{@{}c@{\,\,\,\,\,\,} : c@{\,\,\,\,\,\,}  : c@{\,\,\,\,\,\,} : c@{}}{ 
				{A}_h    & {B}_{ u_h} & {B}_{\perfi_h} & {B}_{\intercPi_h} \\
				\hdashline
				{C}_{y_h }  & {D}_{y u_h} & {D}_{y \perfi_h} & {D}_{y\intercPi_h}  \\
				\hdashline
				{C}_{\perfo_h}  & {D}_{\perfo u_h} & 0 & {D}_{\perfo \intercPi_h} \\
				\hdashline
				{C}_{\intercPo_h}   & {D}_{\intercPo u_h} & {D}_{\intercPo \perfi_h} & 0 \\
			}
			\bma{@{}c@{}}{ 
				\xP_h \\
				\hdashline 
				u_h \\
				\hdashline 
				\perfi_h \\
				\hdashline
				\intercPi_h 
			} , \hspace{0.25em} h \in \mathcal{N}, 
		\end{cases} \\
		%
	\end{aligned}
\end{equation}
with the state vector $\xP_h \in \mathbb{R}^{n_{\xP_h}}$, 
the local control input and measublack output, ${u}_h \in \mathbb{R}^{n_{u_h}}$ and ${y}_h \in \mathbb{R}^{n_{y_h}}$, the local exogenous input and performance output, $\perfi_h \in \mathbb{R}^{n_{\perfi_h}}$ and $\perfo_h \in \mathbb{R}^{n_{\perfo_h}}$, and the interconnection signals $\intercPi_h \in \mathbb{R}^{n_{\intercPi_h}}$ and $\intercPo_h \in \mathbb{R}^{n_{\intercPo_h}}$, respectively\footnote{{\color{black} It is worth noticing that modeling the system with the interconnection channel $p$ to $q$ is equivalent to the system model in \eqref{eq:origsys_matrices}.}}. {\color{black} In this paper we assume $q_h = [{x_h}^\top,  {u_h}^\top]^\top$, namely that
 $C_{q_h} = \begin{bmatrix} I\\ 0\end{bmatrix}$, $D_{qu_h} = \begin{bmatrix}
0 & 0 \\
0 & I
\end{bmatrix}$ and $D_{qw_h} = 0$.
Consequently,  $p_h = [{x_h^i}^\top,  {u_h^i}^\top]^\top$,  where ${x_h^i}^\top,{u_h^i}^\top$ are the overall interconnected state   and  control action of the neighboring subsystems $G_j$ of subsystem $G_h$ \footnote{{\color{black}In other words,} $x_h^{i} = \sum_{j:[P]_{h,j}=1}x_j$ and  $u_h^{i} = \sum_{j:[P]_{h,j}=1}u_j$,  according to \eqref{eq:interconn}. }. }

Based on Definition~\ref{def:homo_sys} and equation \eqref{eq:M_homo}, the state-space matrices $A_h, B_{uh}, C_{yh}, B_{wh}, D_{ywh}, C_{zh}, D_{zuh}$ in \eqref{eq:interconn_openloop_h} are the block-diagonal parts of the system matrices in \eqref{eq:origsys_matrices}, and the system matrices belonging to the interconnection channel $C_{qh}, D_{quh}, D_{qwh}, B_{ph}, D_{yph}, D_{zph}$ are composed of the off-block-diagonal subsystem matrices from \eqref{eq:origsys_matrices}, i.e., 
\begin{align}\label{eq:list_hnterconn_openloop_h}
   B_{\intercPi_h} &= [{A^i}, B^i_{u}],  D_{{y\intercPi}_h} = [{C}^i_{y}, {D}^i_{{y u}}], D_{{\perfo \intercPi}_h} = [{C}^i_{\perfo},{D}^i_{{\perfo u}}], \\  
  C_{\intercPo_h} &= C^d_{\intercPo_h}, D_{\intercPo u_h} = [{D}^d_{{\intercPo u}} ,{D}^i_{{\intercPo u}}],  D_{\intercPo w_h} = {D}^d_{{\intercPo \perfi}}.  \notag
\end{align} 
Note that the representations in \eqref{eq:list_hnterconn_openloop_h} is not unique: choosing the interconnection channel is a modeling choice.

\begin{myrem} 
    We will assume $D^d_{zw}$ to be zero since the feedthrough matrix $D^d_{zw}$ is often chosen to be zero by convention.
\end{myrem}

\section{Interconnected Controller Synthesis}\label{sec:problem_formilation}

We consider interconnected static state feedback controllers, whose  interconnection topology is the same as the one of the system, i.e., the pattern matrix of the controller is equal to the one of the system. 
The controller $K$ for the distributed state feedback  $u(t) = K x(t),$ can thus be written as 
\begin{equation}\label{eq:K_kron}
	K = \kronecker{I_N}{K^d} + \kronecker{\pat}{K^i},
\end{equation}
with $K^d$ being the decentralized local controller gain and $K^i$ the interconnected controller gain.

 We will present a method to synthesize a controller as in \eqref{eq:K_kron} for homogeneous systems described as in Definition~\ref{def:homo_sys}, that minimizes the $\mathcal{H}_\infty$ norm  of the transfer function of the closed-loop system, from the exogenous input $w$ to the performance output $z$, denoted by $T_{zw}$.

The advantage of designing a distributed controller with the structure given in \eqref{eq:K_kron} lies in the fact that the closed-loop system matrices have the same structure as in \eqref{eq:M_homo}, which will prove to be advantageous in terms of scalability of the controller synthesis equations. {\color{black} If we apply such a state feedback law to the system in \eqref{eq:interconn_openloop_h} in closed-loop, we obtain}:
\vspace{-0.1cm}
\begin{equation}\label{eq:clp}
   \small \bma{@{}c@{}}{  
\dot{x}  \\ q \\  z} =
 \small \bma{@{}c@{\,\,\,\,\,\,}c@{\,\,\,\,\,\,}c@{}}{\mathcal{A} & \mathcal{B}_1 & \mathcal{B}_2 \\ 
 \mathcal{C}_1  & \mathcal{D}_{11} & \mathcal{D}_{12} \\
  \mathcal{C}_2  & \mathcal{D}_{21} & \mathcal{D}_{22}}
   \!\! \!\!
  \small \bma{@{}c@{}}{  
x  \\ p \\  w}, \quad p = \mathcal{P} q,
\end{equation}
where the meanings and sizes of the variables $x$, $q$ and $z$ are the same as in the previous section and
	\begin{align}\label{eq:cl_blkdiag}
	   &\mathcal{A} = \kronecker{I_N}{(A^d+[B^d_{u},B^i_{u}]\hat{K})} , \quad \mathcal{B}_1 = \kronecker{I_N}{[A^i,B^d_{u}]}, \notag \\ &\mathcal{B}_2 =\kronecker{I_N}{ B^d_{{\perfi}}},
	   \quad \mathcal{C}_1 = \kronecker{I_N}{(C^d_{{q}}+[D^d_{{qu}},D^i_{{qu}}]\hat{K})},\notag \\
	   & \mathcal{D}_{11} = 0, \quad \mathcal{D}_{12} = \kronecker{I_N}{D^d_{{q\perfi}}}\\
	   &\mathcal{C}_{2} = \kronecker{I_N}{(C^d_{{z}}+[D^d_{{\perfo u}},D^i_{{\perfo u}}]\hat{K})}, \notag \\ &\mathcal{D}_{21} = \kronecker{I_N}{[C^i_{{z}},D^d_{{zu}}]}, \quad \mathcal{D}_{22} = 0, \notag
	\end{align}
 with $\hat{K} = \begin{bmatrix}K^{d\top}, K^{i\top}\end{bmatrix}^\top$.
In this paper we will assume the system \eqref{eq:origsys_matrices} to be stabilizable and the system \eqref{eq:clp} to be well-posed, see \cite{Scherer2001}.
In the following, we present a novel method  to synthesize the controller, which is based on the FBSP presented in \cite{Scherer2001}. It is an extended formulation in which an additional set of variables is introduced according to the \textit{Extended Bounded Real Lemma} (EBRL), presented in \cite{Farhoodi2008}. 
This allows us to decouple the Lyapunov matrix from the other optimization matrices in the EFBSP, thus relaxing the block-diagonal structure of the Lyapunov matrix. This will prove to reduce the conservatism with respect to the standard FBSP used in \cite{massioni2010, Stuerz}, where block-diagonal Lyapunov and multiplier matrices are assumed.
\section{Extended Full Block S-Procedure}\label{sec:FBSP}
\subsection{Extended Full Block S-Procedure in Primal Form }\label{sec:prima_FBSP}
In this section we propose a novel method for the synthesis of a distributed state-feedback controller for the continuous time closed loop system described as in \eqref{eq:clp} and \eqref{eq:cl_blkdiag}. 
The proposed method is an extension of the state-feedback synthesis problem proposed in \cite{Scherer2001}, the FBSP that we recall in the following theorem. 
%
%
%
\begin{mythe}\textit{Full Block S-Procedure} {\cite{Scherer2001}}
\label{the:sproc}
Given an asymptotically stable continuous-time LTI system described as in \eqref{eq:clp}, then the system has an $\mathcal{L}_2$-gain from $w$ to $z$ smaller than $\gamma$ if and only if there exist matrices $\mathcal{X} = \mathcal{X}^\trans > 0$, $R= R^\trans$, $Q= Q^\trans$ and $S$ of appropriate dimensions such that 
\begin{align}
\label{eq:primalLMIs_mult}
 \bma{@{}c@{}}{ 
\star
}^T 
\small \bma{@{}c@{\,\,\,\,\,} c@{}}{ 
Q   & S \\
S^\trans & R
}^\trans
\small \bma{@{}c@{}}{ 
\mathcal{P} \\
I } 
& > 0  , \\
 \bma{@{}c@{}}{ 
\star
}^\trans 
\small \bma{@{}c@{\,\,\,\,\,} c@{\,\,\,\,\,} : c@{\,}  c@{\,} : c@{\,\,\,\,\,}  c@{\,\,\,\,\,}}{ 
0 & \mathcal{X} & 0 & 0 & 0 & 0 \\
\mathcal{X} & 0 & 0 & 0 & 0 & 0 \\
\hdashline
0 & 0 &  Q & S & 0 & 0  \\
0 & 0  & S^\trans & R& 0 & 0\\
\hdashline
0 & 0  & 0 & 0 & - \gamma^2 I & 0 \\
0 & 0 & 0  & 0 & 0 &  I
} 
\small \bma{@{}c@{\,\,\,\,\,\,} c@{\,\,\,\,\,\,}  c@{}}{ 
I & 0 & 0 \\
\mathcal{A} & \mathcal{B}_1 & \mathcal{B}_2 \\
\hdashline
 0 & I & 0 \\
\mathcal{C}_1 & \mathcal{D}_{11} & \mathcal{D}_{12} \\
\hdashline
 0 & 0 & I \\
\mathcal{C}_2 & \mathcal{D}_{21} & \mathcal{D}_{22}
} 
&< 0 .
\label{eq:primalLMIs}
\end{align}
\end{mythe}
\vspace{0.1cm}
\if01
We show in the following proposition how the conditions in 
Theorem~\ref{the:sproc} applied to the interconnected system in \eqref{eq:closed_loop_aug} decompose into conditions of the size of the subsystems by appropriate structural assumptions on the multipliers $Q$, $R$ and $S$ and on the Lyapunov matrix $\mathcal{X}$.  
We introduce the system matrices of subsystem $i$ in \eqref{eq:closed_loop_aug} as 
\begin{gather}
\begin{aligned}
\label{eq:closed_loop_aug_h}
\hspace{-0.18cm} 
\small \bma{@{}c@{\,\,\,\,\,\,}c@{\,\,\,\,\,\,} c@{}}{  
\mathcal{A}_{i}     & \mathcal{B}_{1,i}  & \mathcal{B}_{2,i}   \\ 
\mathcal{C}_{1,i}     & \mathcal{D}_{11,i}  & \mathcal{D}_{12,i}   \\ 
\mathcal{C}_{2,i}     & \mathcal{D}_{21,i}  & \mathcal{D}_{22,i}   
 }
 \!\! &=   \!\! \small \bma{@{}c@{\,\,\,\,\,\,}c@{\,\,\,\,\,\,} : c@{\,\,\,\,\,\,} : c@{}}{  
\bd{A}_{\xi i} & \bd{B}_{\xi u i} K_{\xi i}        & \bd{B}_{\xi \bar{w} i} & \bd{B}_{p i}  \\ 
 \Gamma_h \bd{C}_{y \xi i} & \Phi_h              & 0 & 0 \\
\hdashline
{C}_{\bar{z} \xi ii} & \bd{D}_{\bar{z} u i} K_{\xi i} & \bd{D}_{\bar{z} \bar{w} i} & \bd{D}_{\bar{z} p i}  \\ 
 \hdashline 
\bd{C}_{q i} & \bd{D}_{qu i} K_{\xi i} & \bd{D}_{q \bar{w} i} & 0   }.
\end{aligned}
\end{gather}
\fi 
%
%
%

Equations \eqref{eq:primalLMIs_mult} and \eqref{eq:primalLMIs} are known in the literature as \textit{multiplier condition} and \textit{nominal condition}, respectively.
In this work, the FBSP will be exploited for the design of distributed controllers \cite{massioni2010}. A decomposed scalable distributed control design can be obtained by imposing on the Lyapunov matrix $X$ and the multiplier matrices $Q,R,S$ a block-diagonal structure, which in general introduces conservatism. 
In order to reduce this conservatism, we propose an extended formulation of the FBSP applied to distributed control  presented in \cite{massioni2010} which we refer to as EFBSP. 
%
%
%
We first present the primal formulation, and then dualize it for implementation purposes. The primal formulation is given in the following theorem. 
\begin{mythe}\textit{Extended FBSP in Primal Form}\label{the:Primal_Extended_BRL_FBSP}
	The well-posed rational interconnected system in \eqref{eq:clp} 
	has the $\mathcal{L}_2$-gain from $w$ to $z$ smaller than $\gamma$  if and only if  
	 there exist matrices $X = X^\top > 0$, $F$ non-singular, and $Q = Q^\top$, $R=R^\top$, $S$, such that
		\begin{equation}\label{eq:multipliers_primal}
		\bma{@{}c@{}}{\star}^\top 
		\bma{@{}c@{\,\,\,\,\,\,}c@{}}{
			Q & S  \\ 
			S^\top & R  
		}
		\bma{@{}c@{}}{
			\mathcal{P}   \\ 	
			I 
		} > 0, 
		\end{equation}
%
        \begin{equation}\label{eq:primal_Extended_FBSP}
		\!\!\!\!\!\!\!\!\!\!\!\!\!\!\!\!
		\bma{@{}c@{}}{\star}^\top 
		\!\!\!
		\footnotesize \bma{@{}c@{\,} c@{\!}  @{\,} c @{\!\!\!\!\!\!\!\!} @{\,\,\,\,\,\,\,\,\,\,\,\,}c @{\!\!\!\!\!\!\!\!}: @{\,\,\,\,\,\,\,\,\,\,\,\,\,\,}c@{\,\,\,\,\,\,} c@{\,\,\,\,\,\,} : @{\,\,}c@{\,\,\,\,\,\,}c@{}}{
			0 & X-F & F &0& 0 & 0 & 0 & 0\\ 
			X-F^\top & -F-F^\top &0 & F & 0 & 0 & 0 & 0\\
			F^\top & 0 &0 & 0 & 0 & 0 & 0 & 0\\ 	
			0 & F^\top &0 & 0 & 0 & 0 & 0 & 0\\ 
			\hdashline
			0 & 0 & 0 &0 & Q & S & 0 & 0   \\
			0 & 0 & 0&0  & S^\top & R  & 0 & 0 \\ 
			\hdashline
			0 & 0 & 0 &0& 0 & 0 & -{I\gamma^2} & 0   \\
			0 & 0 & 0 & 0&0 & 0 & 0 & I   
		}
		\!\! 
		\small \bma{@{}c@{\,\,\,\,\,\,}c@{\,\,\,\,\,\,} : 
			c@{} :c@{} }{
			I & 0 & 0 & 0  \\ 	
			0 & I & 0 & 0  \\ 			
			\mathcal{A} & 0 &  \mathcal{B}_1 &  \mathcal{B}_2  \\ 
			\mathcal{A} & 0 &  \mathcal{B}_1 &  \mathcal{B}_2  \\ 
			\hdashline
			0 & 0 & I & 0   \\
			\mathcal{C}_1 & 0 & \mathcal{D}_{11} & \mathcal{D}_{12}  \\ 
			\hdashline
			0 & 0 & 0 & I  \\			
			\mathcal{C}_2 & 0 & \mathcal{D}_{21} & \mathcal{D}_{22}   
		}
		\!\! < \! 0
		\end{equation} 
\end{mythe}
\vspace{-0.5cm}
\begin{mypro} 
	 Consider the extended robust performance conditions expressed in Theorem 2 of \cite{Farhoodi2008}. 
We note that these conditions can equivalently be expressed as follows:
\vspace{-0.cm}
	 There exist $X = X^\top > 0$ and $F$ non-singular, such that 

		\begin{equation}\label{eq:primal_Extended_BRL}
		\!\!\!\!\!\!\!\!\!
		\bma{@{}c@{}}{\star}^\top 
		\footnotesize\bma{@{}c@{\,} c@{\!}  @{\,} c @{\!\!\!\!\!\!\!\!\!} @{\!\!\!\!\!\!\!\!\!\!\!\!\!\!\!\!\!\!} c@{\,} : @{\,\,\,\,\,\,\,\,\,\,\,\,\,} c @{\,\,\,\,\,\,\,}c@{}}{
			0 & X-F & 0 & F & 0 & 0 \\ 
			X-F^\top & -F-F^\top &F &0 & 0 & 0  \\
			F^\top &0 &0& 0 & 0 & 0\\ 	
			0& F^\top  &0& 0 & 0 & 0\\ 	
			\hdashline
			0 & 0 & 0 & 0&-{\gamma^2 I} & 0  \\
			0 & 0 & 0 & 0  &0& I  
		}
		\small \bma{@{}c@{\,\,\,\,\,\,}c@{\,\,\,\,\,\,}:c@{}}{
			I & 0 & 0  \\ 	
			0 & I & 0  \\ 
			\mathcal{A}(\mathcal{P}) & 0 &  \mathcal{B}_1(\mathcal{P}) \\ 	
			\mathcal{A}(\mathcal{P}) & 0 &  \mathcal{B}_1(\mathcal{P}) \\ 
			\hdashline
			0 & 0 & I   \\
			\mathcal{C}_1(\mathcal{P}) & 0 & \mathcal{D}_{11}(\mathcal{P})  
		} 
		< 0, 
		\end{equation}
where the matrices $\mathcal{A}(\mathcal{P})$, $\mathcal{B}_1(\mathcal{P})$, $\mathcal{C}_1(\mathcal{P})$ and $\mathcal{D}_{11}(\mathcal{P})$ are the closed-loop system matrices of 
\begin{equation}\label{eq:clBRL}
	    \begin{cases}
	    \dot{x}(t) &= \mathcal{A}(\mathcal{P})x(t) +  \mathcal{B}_1(\mathcal{P})w(t) \\ 	
		z(t) &=	\mathcal{C}_1(\mathcal{P})x(t)+ \mathcal{D}_{11}(\mathcal{P}) w(t), 
	    \end{cases}
	\end{equation}
	defined as 
$\mathcal{A}(\mathcal{P}) =\kronecker{I_N}{(A^d+B^d_{u}K^d)}+\kronecker{P}{(A^i+B^d_{u}K^i)}$, and the other state-space matrices are defined in the same way, in accordance with \eqref{eq:M_homo}, $\mathcal{B}_1(\mathcal{P}) =\kronecker{I_N}{B^d_{w}}$, $\mathcal{C}_1(\mathcal{P})=\kronecker{I_N}{(C^d+D^d_{u}K^d)}+\kronecker{P}{(C^i+D^d_{u}K^i)}$, $\mathcal{D}_{11}(\mathcal{P})=\kronecker{I_N}{D^d_{w}}$.

	Now we prove the equivalence between \eqref{eq:primal_Extended_BRL} 
	and \eqref{eq:multipliers_primal}- \eqref{eq:primal_Extended_FBSP}.
	To do so, we make use of the FBSP in Theorem~\ref{the:FBSP} in the Appendix.  
	Equation~\eqref{eq:primal_Extended_BRL}, with the following definitions 
	\begin{equation} \label{matrices_EFBSP}
	\begin{aligned} 
	%
	x &= \small\bma{@{}c@{}}{x \\ \dot{x}\\ \dot{x} \\ \dot{x}\\ p \\ q \\ w \\ z}, \quad
	%
    W = \small\bma{@{}c@{\,\,\,\,\,\,}c@{\,\,\,\,\,\,}c@{\,\,\,\,\,\,}c@{\,\,\,\,\,\,}:c@{\,\,\,\,\,\,}c@{}}{
		0 & X-F & F & 0 & 0 & 0 \\ X-F & -F-F^\top &0& F & 0 & 0 \\ F^\top & 0&0 & 0 & 0 & 0 \\
		0 &F^\top &0 & 0 & 0 & 0 \\
		\hdashline 0 & 0 & 0 & 0& - \gamma^2 I & 0 \\ 0& 0 & 0 & 0 & 0 & I}, \\
	V &= \small\bma{@{}c@{\,\,\,\,}c@{\,\,\,\,}c@{\,\,\,\,}c@{\,\,\,\,}:c@{\,\,\,\,}c@{\,\,\,\,}:c@{\,\,\,\,}c@{}}{ 0 & 0 &0& 0 & I & 0 & 0 & 0 \\ 0 & 0 &0& 0 & 0 & I & 0 & 0}, ~
	U = \small\bma{@{}c@{\,\,\,\,\,\,}c@{\,\,\,\,\,\,}c@{\,\,\,\,\,\,}c@{\,\,\,\,\,\,}:c@{\,\,\,\,\,\,}c@{\,\,\,\,\,\,}:c@{\,\,\,\,\,\,}c@{}}{I & 0 &0 & 0 & 0 & 0 & 0 & 0\\ 0 & I &0& 0 & 0 & 0 & 0 & 0\\ 0 & 0 & I&0 & 0 & 0 & 0 & 0\\0 & 0 & 0 & I & 0 & 0 & 0& 0\\\hdashline 0 & 0 & 0 & 0 & 0&0 & I & 0 \\ 0 & 0 & 0 &0 & 0 & 0 & 0 & I}, 
	\end{aligned}
	\end{equation}
can be written as condition (i) of the FBSP in Theorem~\ref{the:FBSP}. This shows that  
	\eqref{eq:primal_Extended_BRL}  holds if and only if both \eqref{eq:multipliers_primal} and \eqref{eq:primal_Extended_FBSP} in Theorem~\ref{the:Primal_Extended_BRL_FBSP} hold, as a result of the FBSP in Theorem~\ref{the:FBSP}\footnote{{\color{black}The advantage of introducing an interconnection channel lies in the reduction of the controller synthesis computational effort that is obtained by imposing a particular structure on the multipliers $Q,R,S$ and decomposing the synthesis equations at the price of introducing some conservatism.}}.  
\end{mypro}

\subsection{Extended FBSP in Dual Form}\label{sec:dual_EBRP}

\begin{mythe}\textit{Extended FBSP in Dual Form} \label{the:dualEFBSP}
	The well-posed rational interconnected system in \eqref{eq:clp} 
	has the $\mathcal{L}_2$-gain from $w$ to $z$ smaller than $\gamma$ if and only if the following statement holds. 
There exists ${Y} =\tilde{F}Y^{-1}\tilde{F}^\top ={Y}^\top > 0$, with $\tilde{F}$ non-singular and multiplier matrices $\tilde{Q}=\tilde{Q}^\top, \tilde{R}=\tilde{R}^\top, \tilde{S}$, 
such that \eqref{eq:multipliers_dual} and \eqref{dualEFBSP} hold.

		\begin{figure*}
			\begin{equation}\label{eq:multipliers_dual}
		\bma{@{}c@{}}{\star}^\top 
		\bma{@{}c@{\,\,\,\,\,\,}c@{}}{
			\tilde{Q} & \tilde{S}  \\ 
			\tilde{S}^\top & \tilde{R}  
		}
		\bma{@{}c@{}}{
			 I  \\ 	
			 \mathcal{P}
		} < 0, 
		\end{equation}
		\begin{equation}\label{dualEFBSP}
			\small \bma{@{}c@{}}{\star}^\top 
			\small \bma{@{}c@{} @{\,\,\,}c@{\,\,\,}c@{\,\,\,}c@{\,\,\,} : @{\,\,\,}c@{\,\,\,\,} c@{\,\,\,\,} : @{\,\,\,\,}c@{\,\,\,}c@{\,\,\,}}{
				0 & 0 & \textcolor{black}{\tilde{F}} &0 & 0 & 0 & 0 & 0\\ 
				0 & 0  &0 & \textcolor{black}{\tilde{F}} & 0 & 0 & 0 & 0\\ 
				\textcolor{black}{\tilde{F}^\top} & 0 &0 & \textcolor{black}{\tilde{F}}-\textcolor{black}{{Y}}  & 0 & 0 & 0 & 0\\ 	
				0& \textcolor{black}{\tilde{F}^\top} & \textcolor{black}{\tilde{F}^\top}-\textcolor{black}{{Y}} & \textcolor{black}{\tilde{F}}+\textcolor{black}{\tilde{F}^\top} & 0 & 0 & 0 & 0\\ 
				\hdashline
				0 & 0 & 0 &0 & \tilde{Q} & \tilde{S} & 0 & 0  \\
				0 & 0 & 0 &0 & \tilde{S}^\top & \tilde{R}  & 0 & 0 \\ 
				\hdashline
				0 & 0 & 0 & 0& 0 & 0 & -\frac{1}{\gamma^2}I & 0   \\
				0 & 0 & 0 & 0 &0 & 0 & 0 & I   
			}
			\small \bma{@{}c@{\,\,}@{\,\,}c@{}@{\,\,}:c@{}:c@{} }{
				-\mathcal{A}^\top & 0 &  -\mathcal{C}_1^\top &  -\mathcal{C}_2^\top  \\ 
				-\mathcal{A}^\top & 0 &  -\mathcal{C}_1^\top &  -\mathcal{C}_2^\top  \\ 
				I & 0 & 0 & 0  \\ 	
				0 & I & 0 & 0  \\ 	
				\hdashline
				-\mathcal{B}_1^\top & 0 & -\mathcal{D}_{11}^\top & -\mathcal{D}_{12}^\top  \\ 
				0 & 0 & I & 0   \\
				\hdashline
				-\mathcal{B}_2^\top & 0 & -\mathcal{D}_{21}^\top & -\mathcal{D}_{22}^\top  \\ 
				0 & 0 & 0 & I  
			}>0
		\end{equation} 
		\hrule
		\end{figure*}
	
\end{mythe}
\begin{mypro}
     We start from the primal formulation of the EFBSP in Theorem~\ref{the:Primal_Extended_BRL_FBSP} and we exploit the Dualization Lemma reported in Lemma 4.8 from \cite{LMI} 
     in which the middle matrices in \eqref{eq:primal_Extended_FBSP} and in \eqref{dualEFBSP} play the same roles as $P$ and $P^{-1}$ in Lemma 4.8 from \cite{LMI},   respectively. Then, by considering the dual system with respect to the system in \eqref{eq:clBRL}, i.e.
     \begin{equation}\label{eq:clBRL_dual}
	    \begin{cases}
	    \dot{x}(t) &= \mathcal{A}^\top(\mathcal{P})x(t) +  \mathcal{C}_1^\top(\mathcal{P})w(t) \\ 	
		z(t) &=	\mathcal{B}_1^\top(\mathcal{P})x(t)+ \mathcal{D}_{11}^\top(\mathcal{P}) w(t), 
	    \end{cases}
	\end{equation}
     the dual formulation as in  \eqref{eq:multipliers_dual}-\eqref{dualEFBSP} is obtained and this concludes the proof.
\end{mypro}

\normalsize

\section{Decomposed Extended FBSP} \label{sec:EFBSP_decomp}

In this section we present the main result of this paper, which is a decomposed EFBSP for scalable interconnected controller synthesis, 
given in the following theorem. 


\begin{mythe}\label{the:decomposed_dualEFBSP}
    \textit{Decomposed EFBSP in Dual Form:} 
The well-posed rational interconnected system in \eqref{eq:clBRL}, whose interconnection matrix $\mathcal{P}=\kronecker{P}{I_N}$ is symmetric, has   $\mathcal{L}_2$-gain from $w$ to $z$ smaller than $\gamma$ if
	there exist $Y^d$, $Y^i$, $\tilde{F}^d$ non-singular, 
	such that ${Y} =\tilde{F}Y^{-1}\tilde{F}^\top ={Y}^\top > 0$, with $Y = \kronecker{I}{Y^d} + \kronecker{P}{Y^i}$,  $\tilde{F} = \kronecker{I_N}{\tilde{F}^d}$  and there exist matrices  $\tilde{Q}^i$, $\tilde{Q}^d$,   $\tilde{R}^i$, $\tilde{R}^d$,  $\tilde{S}^d$, $\tilde{S}^i$, with $\tilde{Q}^d=\tilde{Q}^{d\top}$, 
	$\tilde{R}^d=\tilde{R}^{d\top}$,
	and $M^d$, 
	such that \eqref{eq:multipliers_dual_decom} and \eqref{equivalent_dualFBSP_decomp} hold, where $\lambda_h \in \mathrm{spec}({{P}})$,  $h=1,\dots,N$, are the eigenvalues of $P$. 
		\begin{figure*}
			\begin{equation}\label{eq:multipliers_dual_decom}
	\tilde{Q}^d+\lambda_h(\tilde{Q}^i-\tilde{S}^{d\top}-\tilde{S}^d)+\lambda_h^2(-\tilde{S}^{i\top}-\tilde{S}^i+\tilde{R}^d)+\lambda_h^{3}\tilde{R}^i
	<0   \quad \forall \lambda_h \in \mathrm{spec}({P}),
		\end{equation}
		\begin{equation}\label{equivalent_dualFBSP_decomp}
		\begin{aligned}
	       \small\bma{@{}c@{}}{\star}^\top 
	       	\scriptsize \bma{@{}c@{\!\!\!} @{\,\,\,\,\,\,\,\,\,\,}c@{\,\,\,\,}c@{\,\,\,\,}c@{\!\!\!} : @{\,\,\,\,\,}c@{\,\,\,} c@{\,\,\,} : @{\,\,}c@{\,\,\,}c@{}}{
				0 & 0 & I &0 & 0 & 0 & 0 & 0\\ 
				0 & 0  &0 & I & 0 & 0 & 0 & 0\\ 
				I & 0 &0 & \tilde{F}^d-{Y}^d-\lambda_h{Y}^i  & 0 & 0 & 0 & 0\\ 	
				0& I & \tilde{F}^{d\top}-{Y}^d-\lambda_h{Y}^i  & \tilde{F}^d+\tilde{F}^{d\top} & 0 & 0 & 0 & 0\\ 
				\hdashline
				0 & 0 & 0 &0 & \tilde{Q}^d+\lambda_h\tilde{Q}^i & \tilde{S}^d+\lambda_h\tilde{S}^i & 0 & 0  \\
				0 & 0 & 0 &0 & \tilde{S}^{d\top}+\lambda_h\tilde{S}^{i\top} & \tilde{R}^d+\lambda_h\tilde{R}^i  & 0 & 0 \\ 
				\hdashline
				0 & 0 & 0 & 0& 0 & 0 & -\frac{1}{\gamma^2}I & 0   \\
				0 & 0 & 0 & 0 &0 & 0 & 0 & I   
			}
			\footnotesize \bma{@{\,}c@{\,\,\,\,\,}c@{\,\,\,\,\,}:c@{\,\,\,\,\,}:c@{\,\,\,\,\,} }{
				-(A_h\tilde{F}^d+B_{u_h}M^d)^\top & 0 & -({C}_{q_h}\tilde{F}^d+D_{qu_h}M^d)^\top  &  -({C}_{z_h}\tilde{F}^d+D_{zu_h}M^d)^\top \\ 
					-(A_{i}\tilde{F}^d+B_{u_h}M^d)^\top & 0 & -({C}_{q_h}\tilde{F}^d+D_{qu_h}M^d)^\top  &  -({C}_{z_h}\tilde{F}^d+D_{zu_h}M^d)^\top  \\ 
				I & 0 & 0 & 0  \\ 	
				0 & I & 0 & 0  \\ 	
				\hdashline
				-\mathcal{B}_{1_h}^\top & 0 & -\mathcal{D}_{{11}_h}^\top & -\mathcal{D}_{{12}_h}^\top  \\ 
				0 & 0 & I & 0   \\
				\hdashline
				-\mathcal{B}_{2_h}^\top & 0 & -\mathcal{D}_{{21}_h}^\top & -\mathcal{D}_{{22}_h}^\top  \\ 
				0 & 0 & 0 & I  
			}&>0 , \\
			\forall {h=1,\dots,N}&
		\end{aligned}
		\end{equation}
		\hrule
		\end{figure*}
%
\end{mythe}
\begin{mypro}
Consider \eqref{eq:multipliers_dual} and \eqref{dualEFBSP} in Theorem~\ref{the:dualEFBSP},  where the closed-loop matrices of the interconnected system under the distributed controller are as in \eqref{eq:cl_blkdiag}.
 We perform the following variable substitution in order to convexify \eqref{dualEFBSP} 
 \begin{equation}\label{eq:MK}
     M = \hat{K}\tilde{F}=\begin{bmatrix}K^d\\ K^i\end{bmatrix}\tilde{F}
 \end{equation}
leading to the closed-loop matrices   in \eqref{eq:cl_blkdiag}. 
We note that these closed-loop matrices all have a block-diagonal structure. 
 Next, choosing the multiplier matrices in \eqref{eq:multipliers_dual} and \eqref{dualEFBSP} structured as 
\begin{equation}\label{eq:QRS_kron}
	\begin{aligned}
		\tilde{Q} = \kronecker{I}{\tilde{Q}^d} + \kronecker{P}{\tilde{Q}^i},& \quad
		\tilde{S} = \kronecker{I}{\tilde{S}^d} + \kronecker{P}{\tilde{S}^i} , \\
		\tilde{R} = \kronecker{I}{\tilde{R}^d} &+ \kronecker{P}{\tilde{R}^i},  
	\end{aligned}
\end{equation}
we apply the following algebraic transformation. Let $Z$ be an orthonormal matrix such that $Z^{\top}PZ = \Lambda$, where $\Lambda$ is the diagonal matrix having the eigenvalues of $P$ as diagonal entries, i,e. $\Lambda = diag_{i=1}^N\{\lambda_h\}$. Then, we apply the algebraic transformations 
\begin{align}\label{eq:trafo}
    \hat{E}_{MC} &= \hat{Z}^{\top} \,  E_{MC} \, \hat{Z} \notag \\
   \hat{E}_{NC,1}^\top  \hat{E}_{NC,2}  \hat{E}_{NC,1}&=\hat{H}^{\top} \, E_{NC,1}^\top \, \hat{T}  \, \hat{T}^{\top} \,  E_{NC,2} \, \hat{T}  \, \hat{T}^{\top} \, E_{NC,1} \, \hat{H},
\end{align}
where $E_{MC}$ is the multiplier condition in \eqref{eq:multipliers_dual}, and $E_{NC,1}$ and $E_{NC,2}$ are the matrices that form the nominal condition $E_{NC,1}^\top \, E_{NC,2} \, E_{NC,1}$ in  \eqref{dualEFBSP}, respectively. 
The transformation in \eqref{eq:trafo} is defined by the matrices $\hat{Z}:=Z\otimes I_{z_1}$, $\hat{H} = diag_{i=1}^3\{\hat{H}_i\}$, with $\hat{H}_i := Z\otimes I_{h_i}$, and $\hat{T} = diag_{i=1}^3\{\hat{T}_i\}$ with $\hat{T}_h := Z\otimes I_{t_i}$, $i = 1,2,3$, with $z_1,h_1,h_2,h_3,t_1,t_2,t_3$ being the dimensions of the identity matrices such that the block-partitions of the transformation matrices $\hat{Z}$, $\hat{H}$ and $\hat{T}$ matches the block-partitions of the matrices $E_{MC}$, $E_{NC,1}$ and $E_{NC,2}$ as indicated in \eqref{eq:multipliers_dual} and \eqref{dualEFBSP}, respectively, to which they are applied.

After some basic algebraic manipulations and exploiting the properties of the Kronecker product, it can be seen that the block-diagonal blocks are only multiplied with identity matrices, and that the matrix $P$ is diagonalized, resulting in  LMIs \eqref{eq:multipliers_dual_decom} and \eqref{equivalent_dualFBSP_decomp}.
\end{mypro}






In comparison with the decomposed controller synthesis approach in \cite{massioni2010}, where the block-diagonal structure needs to be imposed on the multiplier matrices $\tilde{Q},\tilde{R},\tilde{S}$ and on the Lyapunov matrix $Y$ in order to decompose the synthesis conditions in the FBSP, imposing the structure in \eqref{eq:QRS_kron} in the EFBSP allows us to obtain the same degree of decomposition 
and thus scalability, i.e., to decompose the synthesis conditions into small matrix inequalities in the size of the individual subsystems, with the advantage of introducing less conservatism. This is achieved by exploiting the diagonalizability of the matrix $P$. 
Moreover, since $P$ is symmetric, the matrix $Z$ that diagonalizes $P$ is orthonormal and consequently the value of $\gamma$ obtained from \eqref{eq:multipliers_dual_decom}-\eqref{equivalent_dualFBSP_decomp} does not differ from the one for the original system. 
More details related to the case in which $P$ is not symmetric are discussed in \cite{massioni2}.


\begin{myrem}
{\color{black} The computational time needed to solve the LMIs \eqref{eq:multipliers_dual_decom}-\eqref{equivalent_dualFBSP_decomp} scales linearly, while the one needed to solve \eqref{eq:multipliers_dual}-\eqref{dualEFBSP} scales polynomially with $N$. Moreover, }
at the price of introducing some conservatism, by imposing the additional constraints $R^i = 0$ and  $-\tilde{S}^{i\top}-\tilde{S}^i+\tilde{R}^d> 0$ in Theorem~\ref{the:decomposed_dualEFBSP}, the constraints in \eqref{eq:multipliers_dual_decom} become quadratic and convex in the variables $\lambda_h$. Therefore one can simply solve it only for the constraints corresponding to the largest and smallest eigenvalues of $P$. 
\end{myrem}
\begin{myrem}

    The structure imposed on the optimization variables $Y, \tilde{Q}, \tilde{R}, \tilde{S}$ involved in the EFBSP can also be chosen as $Y = \kronecker{I_N}{Y^d}+\kronecker{P}{Y^i}+\kronecker{P_1}{Y^i_{1}}+ \dots +\kronecker{P_r}{Y^i_{r}}$
    and the same for $ \tilde{Q}, \tilde{R}, \tilde{S}$, in \eqref{dualEFBSP} and \eqref{eq:multipliers_dual}, where 
    $P, P_1, \dots P_r$ are simultaneously diagonalizable pattern matrices, see \cite{sim_diag} and \cite{massioni2} for more details about simultaneously diagonalizable matrices and their applications. 
    This approach can further reduce conservatism since more non-zero entries, i.e., more degrees of freedom for $Y, \tilde{Q}, \tilde{R}, \tilde{S}$, are allowed. 
 Let $P_1$ be the matrix with scalar entries $p_{ij}$, i.e., $P_1=[p_{ij}]_{i,j=1,\dots,N}$. Let us suppose that $P_1$ is the pattern matrix, simultaneously diagonalizable with $P$, with the highest number of non-zero entries, and hence the matrix allowing the less restrictive structure on $Y,\tilde{Q}, \tilde{R}, \tilde{S}$ among all the simultaneously diagonalizable matrices with $P$, $P_1,P_2,\dots, P_r$. $P_1$ may be found as the solution of the following optimization problem:
    \begin{equation}
    P_1 =\max_{\substack{s.t.\\ P_{1}= P_{1}^{T}\\ PP_{1}=P_{1}P\\p_{ii}=0\\ p_{ij}(1-p_{ij})=0}} \sum_{i < j} p_{ij}.
\end{equation}
where the constraint $PP_{1}=P_{1}P$ fully characterizes the simultaneous diagonalizability \cite{sim_diag}.

\end{myrem}
\vspace{-0.25cm}
\section{Example and Simulation Results}\label{sec:sim_res}
\vspace{-0.15cm}


By referring to the description in \eqref{eq:interconn_openloop_h}-\eqref{eq:list_hnterconn_openloop_h},  we consider a system whose subsystems are described by the following matrices:
\vspace{-0.15cm}
\begin{footnotesize}
\begin{align}
A^d &= \begin{bmatrix}0 & -0.2 & 0.8\\ -0.9 & -0.7 & -0.4 \\ -0.9 & 0.5 & -0.6\end{bmatrix},{B}^d_{u} = \begin{bmatrix}0.2 \\ 0.2 \\ -0.1 \end{bmatrix} {B}^d_{\perfi} = \begin{bmatrix}0.5 \\ 0.5 \\ 0.5 \end{bmatrix} \notag \\   
 {A}^i&= \begin{bmatrix}0.2 & 0.1 & -0.1 \\ -0.4 & -0.1 & -1.0 \\ -0.3 & -0.1 & 0.0 \end{bmatrix},{C}^d_{y} = \begin{bmatrix} 1.1 & -2.1 & -1.6 \\ -0.3 & -0.4 & 0.5 \\ 0.7 & -0.8 & 0.3\end{bmatrix},\notag \\  
{C}^d_{{\perfo}} &= \begin{bmatrix} 0.8 & 0.3 & 0.3 \\ -0.1 & 0.1 & -0.1\end{bmatrix}, {D}^d_{{\perfo u}} = \begin{bmatrix}0.1 \\ 0.9 \end{bmatrix},  {C}^i_{{\perfo}} = \begin{bmatrix} 0 & 0 & 0 \\ 0.3 & 0.2 & 0.5 \end{bmatrix}, \notag 
 \notag
\end{align}
\end{footnotesize}
all the others are set to zero. The subsystems are interconnected according to the pattern matrix
\begin{footnotesize}
\begin{equation*}
    P = \begin{bmatrix}
    0 & 1 & 0 & 0 & 1 & 0\\
    1 & 0 & 1 & 1 & 0 & 1 \\
    0 & 1 & 0 & 0 & 1 & 0 \\
    0 & 1 & 0 & 0 & 1 & 0\\
    1 & 0 & 1 & 1 & 0 & 1 \\
   0 & 1 & 0 & 0 & 1 & 0
  \end{bmatrix}.
\end{equation*}
\end{footnotesize}
Moreover, the interconnection variables are defined as $p_h = [x_h^{i\top},u_h^{i\top}]^\top$ where $x_h^i$ and $u_h^i$ are the interconnected state from and control action depending on neighboring subsystem $G_j$ of subsystem $G_h$.
\normalsize

\textbf{Results.} In the following some numerical results are shown in order to illustrate the proposed controller synthesis via decomposed EFBSP in Theorem~\ref{the:decomposed_dualEFBSP}. Table~\ref{tab:results} shows a comparison between the results obtained by applying the FBSP in which all the optimization variables are restricted to be block-diagonal and the results obtained by applying the EFBSP in which the matrix $\tilde{F}$ is restricted to be block-diagonal, while the Lyapunov matrix $Y$ and the multipliers are structured as in \eqref{eq:QRS_kron}. 
The optimization problem has been coded in $\mathtt{Matlab}$ and solved through the non-linear solver $\mathtt{Penbmi}$ from the $\mathtt{Tomlab}$ toolbox. Since in our simulations the solver did not converge, an additive constraint that restricts $\gamma$ to be lower than the value of $\gamma$ obtained in the standard FBSP with all the optimization variables restricted to have block-diagonal structure has been added.

\begin{small}
\begin{center}
 \begin{tabular}{||c| c| c ||}
  \hline
   & FBSP &  EFBSP   \\ [0.5ex] 
 \hline\hline
  &Y blkdiag $\tilde{Q},\tilde{R},\tilde{S}$ blkdiag& $\tilde{F}$ blkdiag, $(Y,\tilde{Q},\tilde{R},\tilde{S})_{\otimes}$  \\ 
 \hline
  $\gamma$ & 10.2533    & 9.2236 \\
 $\mathcal{H_{\infty}}$ & 10.1814 & 8.0553\\
 \hline
\end{tabular}
\captionof{table}{Numerical results}
\label{tab:results}
\end{center}
\end{small}
It is worth saying that, unlike the EFBSP, relaxing the block-diagonal structure of the multiplier matrices in the standard FBSP to the structure in \eqref{eq:QRS_kron} does not bring any improvement in terms of performance and this is one of the reasons why by means of the EFBSP better performance may be reached. 

\printbibliography

\section*{APPENDIX}\label{sec:theory}

\subsection{Full Block S-Procedure}

\begin{mythe} \cite{Scherer2001}\label{the:FBSP} 
Assume $U \in \mathbb{R}^{p\times n}, V \in \mathbb{R}^{q\times n}, W \in \mathbb{R}^{n\times n}$, with $W=W^\top$, and  let $\mathcal{I}$ be a subspace of $\mathbb{R}^n$. Furthermore, let $\mathcal{S}(\rho) \in \mathbb{R}^{q}$ be a subspace that depends continuously on the parameter $\rho$ which varies in the compact path-connected set $\boldsymbol{\rho}$. Let $\mathcal{B}(\rho)$ be a family of subspaces of $\mathbb{R}^n$, defined as 
	$$\mathcal{B}(\rho):= \{x\in\mathcal{I} \vert Vx \in \mathcal{S}(\rho)\}$$ 
	Then the following two statements are equivalent: \\
	(i) $U^\top W U < 0$ on $\mathcal{B}(\rho)$, $\forall \rho \in \boldsymbol{\rho}$. \\
	(ii) There exists $P_e = P_e^\top > 0$ on $\mathcal{S}(\rho)$, $\forall \rho \in \boldsymbol{\rho}$, which satisfies 
	$U^\top W U + V^\top P_e V < 0$ on $\mathcal{I}$.
\end{mythe}

\end{document}

%% file: nomenclature.tex
\def\qed {{
		\parfillskip=0pt 
		\widowpenalty=10000 
		\displaywidowpenalty=10000 
		\finalhyphendemerits=0 
		\leavevmode 
		\unskip 
		\nobreak 
		\hfil 
		\penalty50 
		\hskip.2em 
		\null 
		\hfill 
		$\blacksquare$
		\par}}

\newenvironment{mypro}[1][\hspace{2.3mm} Proof]{\begin{trivlist} 
		\item[\hskip \labelsep {\itshape #1:}]}{\qed \end{trivlist} }
\newtheorem{myrem}{Remark}

\newtheorem{mythe}{Theorem}

\newtheorem{mydef}{Definition}

\newcommand{\bma}[2]       {\left[\begin{array}{#1}#2\end{array}\right]}	
\newcommand{\mrm}[1]       {\mathrm{#1}}

\newcommand{\diagIndtwo}[3]      {\mrm{\diagonal}_{#1}^{#2}\!\left(#3\right)}		
		
\newcommand{\trans}        {\top}																				  

\newcommand{\kronecker}[2] {{#1}\otimes{#2}}
\newcommand{\pat} {P}

\newcommand{\Plant}[0]{G}

\newcommand{\xP}[0]{x}

\newcommand{\dotxP}[0]{\dot{x}}

\newcommand{\bd}[1]   {{#1}^d}

\newcommand{\perfo}[0]{{z}} 
\newcommand{\perfi}[0]{{w}} 
\newcommand{\intercPo}[0]{q} 
\newcommand{\intercPi}[0]{p} 
